# Hybrid Si-GaAs photonic crystal cavity for lasing and bistability

Mohammad Habibur Rahaman[1,2], Chang-Min Lee[1,2], Mustafa Atabey Buyukkaya[1,2], Yuqi Zhao[1,2], Edo Waks[1,2,3,4]

[1]Department of Electrical and Computer Engineering, University of Maryland, College Park, Maryland, USA, 20742.
[2]Institute for Research in Electronics and Applied Physics (IREAP), University of Maryland, College Park, Maryland, USA, 20742.
[3]Department of Physics, University of Maryland College Park, MD, 20742.
[4]Joint Quantum Institute (JQI), University of Maryland, College Park, MD, 20742.
**Corresponding author**, edowaks@umd.edu

**Abstract.** The heterogeneous integration of silicon with III-V materials provides a way to overcome silicon's limited optical properties toward a broad range of photonic applications. Hybrid modes are a promising way to make heterogeneous Si/III-V devices, but it is still unclear how to engineer these modes to make photonic crystal cavities. Herein, using 3D finite-difference time-domain simulation, a hybrid Si-GaAs photonic crystal cavity design enables cavity mode confinement in GaAs without directly patterning that operates at telecom wavelengths. The hybrid cavity consists of a patterned silicon waveguide nanobeam that is evanescently coupled to a GaAs slab with quantum dots. We show that by engineering the hybrid modes, we can control the degree of coupling to the active material, which leads to a tradeoff between cavity quality factor and optical gain and nonlinearity. With this design, we demonstrate a cavity mode in the Si-GaAs heterogeneous region, which enables strong interaction with the quantum dots in the GaAs slab for applications such as low-power-threshold lasing and optical bistability (156 nW and 18.1 µW, respectively). This heterogeneous integration of an active III-V material with silicon via a hybrid cavity design suggests a promising approach for achieving on-chip light generation and low-power nonlinear platforms.

## 1. Introduction

Silicon is a promising candidate for photonic integrated circuits due to its low power consumption, low cost, and high integration density.[1] Additionally, silicon on an insulator provides a high index contrast platform[2,3] that enables the fabrication of extremely compact passive optical devices, such as low loss waveguides[4,5], grating couplers[6,7], mode converters[8], multiplexers[9], and more. However, silicon lacks active optical properties due to its indirect bandgap, which has impeded the development of silicon-based photonic integrated circuits. As an alternative, direct bandgap III-V semiconductors (e.g., GaAs) offer significant active optical properties such as light emission and absorption at telecom wavelengths. In particular, III-V nanophotonic devices can implement low threshold lasers[10,11], amplifiers[11], modulators[12], and highly nonlinear devices[13]. The integration of III-V nanophotonic devices with silicon could pave the way towards ultra-compact low energy nanophotonic and opto-electronic devices. Researchers have increasingly explored the heterogeneous integration of III-V materials with silicon to utilize the advantages of both systems[2,14].

There has been significant progress in hybrid integration of silicon and III-V nanophotonics. Nanophotonic heterogeneous integration has also been achieved in several nanolaser designs.[15–19] In all of these cases, hybrid integration was achieved by placing a III-V nanophotonic device in close proximity to a silicon waveguide and relying on evanescent coupling between the two structures. This approach requires careful placement and alignment of the nanophotonic cavity in the evanescent field of the waveguide in order to achieve efficient coupling. But such active alignment can significantly complicate the design and fabrication of integrated photonic devices.[20] Hybrid photonic modes offer an alternate approach to couple silicon and III-V semiconductors. In this approach, the mode of a silicon waveguide is carefully engineered to hybridize with a III-V semiconductor layer that serves as the active material.[21] Hybrid modes have been used to engineer laser sources,[21,22] amplifiers,[23] and modulators.[24] But all of these works exploit hybrid modes of a silicon waveguide. The use of hybrid modes to engineer nanophotonic devices remains far less explored.

In this letter we propose and analyze a hybrid mode approach to engineer Si-GaAs nanophotonic cavity without the need for careful active alignment that enables low power threshold lasing and optical bistability. Unlike previous hybrid integration approaches where a silicon waveguide is coupled to a GaAs nanophotonic device, the hybrid mode cavity utilizes a patterned silicon cavity structure coupled to a planar GaAs slab that contains the active material. This approach alleviates the need for careful alignment between a silicon photonic waveguide and III-V semiconductor device. We numerically design and characterize the structure using 3D finite-difference time-domain (FDTD) method coupled to a Maxwell-Bloch equation model that accounts for the active material in the GaAs slab.[25,26] We show that the hybrid cavity mode can exhibit a low lasing threshold of just 156 nW, which we attribute to the high spontaneous emission coupling ratio enabled by the low mode volume of the active media in the hybrid cavity.[27] Additionally, under continuous wave operation, we found a low optical bistability threshold of 18.1 µW. This hybrid Si-GaAs cavity design can also be optimized for different operating frequencies and integration with other active materials beyond quantum dots, such as Kerr nonlinear media.[28] This approach could enable simpler more scalable approach to heterogeneous integration of III-V materials with silicon nanophononics.

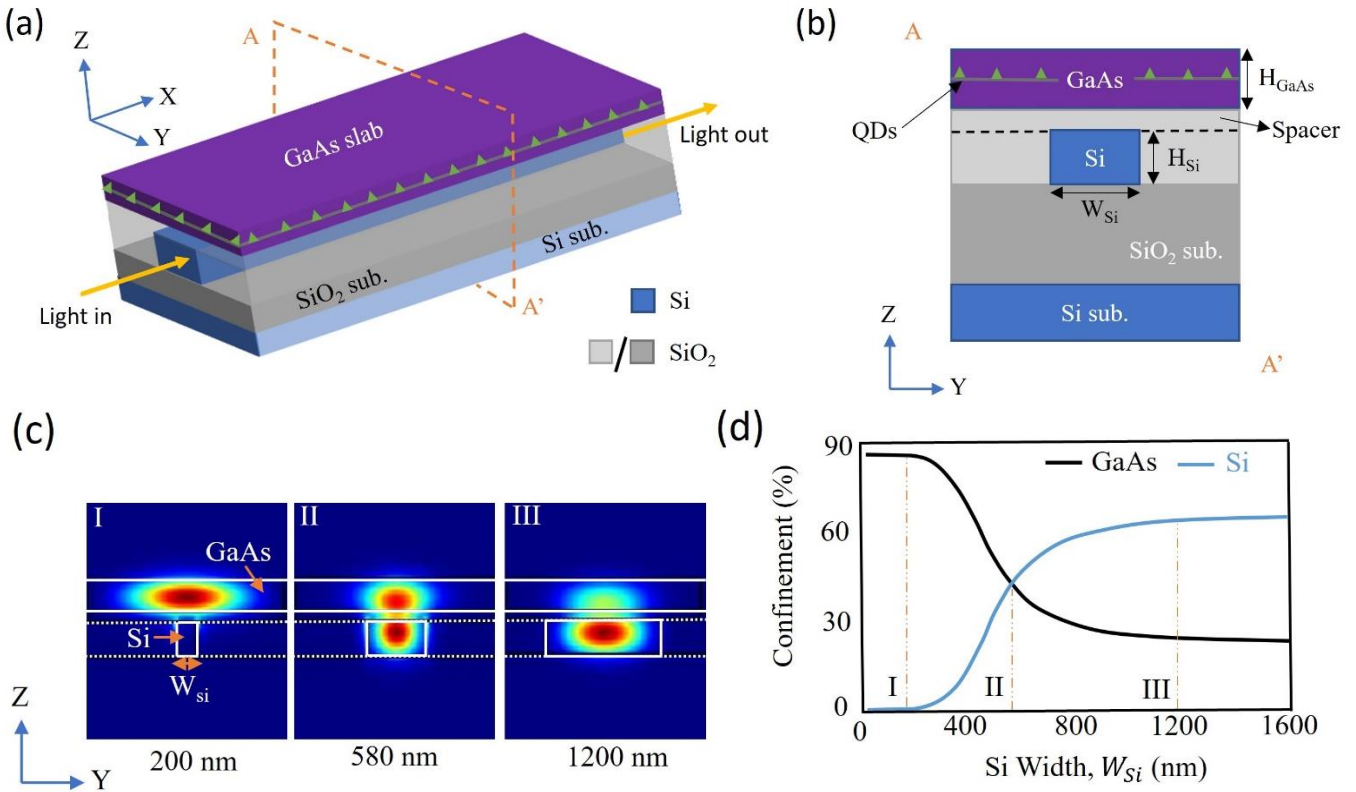

Figure 1. (a) 3D Schematic illustration of the Si-GaAs hybrid cavity, and (b) the corresponding 2D cross-section. (c) The transverse mode profile (|E|) of the hybrid mode for different silicon waveguide widths ($W_{Si}$) of 200 nm, 580 nm, and 1200 nm at a fixed GaAs slab width of 2 µm at an incident light wavelength of 1516.53 nm. The white lines represent the boundaries of the silicon waveguide and GaAs slab. (d) The silicon waveguide width controls the relative confinement of the hybrid mode in the silicon waveguide and GaAs slab region, where positions I, II, and III correspond to different silicon waveguide widths marked in (c).

## 2. Proposed hybrid Si-GaAs cavity design

We first analyze the hybrid mode structure of a silicon waveguide coupled to a GaAs slab. Figure 1a, b illustrates the hybrid structure which is composed of a silicon waveguide with height $H_{Si}$ and width $W_{Si}$. The waveguide is embedded in an $SiO_2$ cladding layer. A GaAs slab of thickness $H_{GaAs}$ and width $W_{GaAs}$ embedded with active material (e.g., quantum wells or quantum dots) is placed on top of the silicon waveguide separated by a spacer with thickness $t$.

We simulate the mode structure of the waveguide by direct eigenmode expansion (ANSYS MODE solutions). We consider the case where $t = 60\ nm$, $H_{Si} = 220\ nm$, $H_{GaAs} = 200\ nm$ and $W_{GaAs} = 2\ \mu m$. Figure 1c shows the transverse mode profile of the waveguide modes for several different values of $w$. Consistent with previous work on hybrid modes[2], as the width of the waveguide increases from 200 nm to 1200 nm the guided mode continuous shifts from being predominantly confined in the GaAs slab to be being confined the silicon waveguide. Figure 1d shows confinement factor in silicon and GaAs region, defined as $\int_{Si\ or\ GaAs}|E|^2\ dA / \int_{total}|E|^2\ dA$, which shows the gradual shift between the two materials. The three vertical dashed lines labeled I, II, and III correspond to the widths plotted in Fig. 1c.

Without the GaAs layer, this device structure would lead to a highly localized cavity mode within the silicon waveguide. But adding the GaAs layer causes the mode to hybridize. The key to designing the cavity mode is to carefully select the width $W_{Si}$. If this width is made to be large, most of the mode will be confined in silicon leading to strong optical confinement but poor overlap with the GaAs active material. In contrast, if $W_{Si}$ is made too small the majority of the hybrid mode will be confined to the GaAs active material where it can experience large gain and nonlinearity. But in this regime the effective index contrast of the photonic crystal will be low which will result in poor optical confinement and low quality factors. The width must therefore be carefully designed to balance out these conflicting requirements.

To design the cavity, we first analyze a periodic photonic crystal waveguide. We choose ellipsoidal cavity with a lattice period $a$ of 380 nm and radii of 140 nm and 130 nm for the major and minor axes of the elliptical cavities, respectively. We select the width of the waveguide to be 580 nm, which creates hybrid mode that is nearly equally confined in both the silicon and GaAs (see Figure 1c). This value provides a good compromise between mode confinement and overlap with the active material. We calculate the photonic band structure by performing three-dimensional Finite Difference Time Domain simulations (ANSYS FDTD solution) with Bloch boundary conditions. Figure 2a shows the optimized photonic band-structure, which features a sufficiently wide photonic bandgap of 8.5 THz that spans within a telecom range of 192 THz to 200.5 THz (grey shaded region).

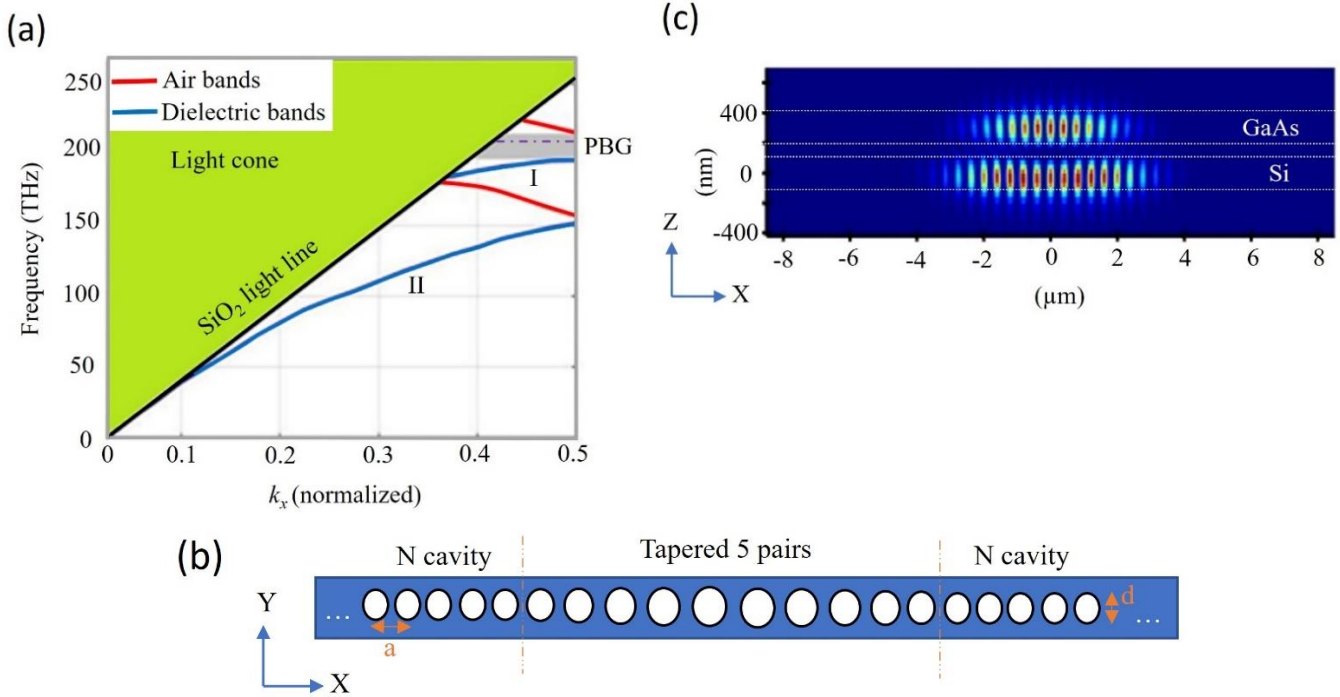

Figure 2. (a) The FDTD calculated band structure of the hybrid Si-GaAs device. The grey shaded area is the photonic bandgap (PBG), which corresponds to the difference between the calculated air bands (red curves) and dielectric bands of the silicon and GaAs (blue curves marked I and II, respectively). Here, $k_x$ is related to wavevector $k = k_x (2\pi/a)$, which is a function of the reciprocal lattice parameter $a$ specified by the Bloch boundary condition. The black line denotes the $SiO_2$ light line, and the horizontal purple dashed line in PBG indicates the cavity resonance frequency. (b) Schematic diagram of the silicon waveguide cavity design, which features a total of 16 cavity pairs, including $N = 11$ ellipsoidal cavities on both sides and 5 centered pairs of tapered ellipsoids as defects. (c) The fundamental mode field ($|E|$) profile in the hybrid Si-GaAs cavity obtained using the 3D FDTD method. The white dashed lines denote the silicon and GaAs layers (220 nm and 200 nm thick, respectively), which are separated by a 60 nm spacer.

In order to create a cavity mode, we introduce a defect into the photonic crystal waveguide. The defect is composed of a taper where we linearly increase both the major and minor axes of the elliptical holes. Figure 2b shows the structure of the defect corresponding to five center pairs of holes taper where we increase the major axes of the ellipses from 140 nm to 150 nm, and the minor axes increase from 130 nm to 140 nm. Figure 2c plots the electric field profile $|E|$ at resonance wavelength 1516.53 nm of this hybrid Si-GaAs cavity, which reveals a fundamental cavity mode that extends in both the silicon waveguide and GaAs slab (Figure 2c). The mode is highly localized in the cavity region and has a substantial overlap with both the silicon and GaAs region. It thus achieves a hybrid localized photonic crystal cavity mode. We calculate the mode volume to be $V_m \sim 2\ (\lambda/n)^3$, and the quality factor to be $Q = 1.53 \times 10^4$.

## 3. Lasing

The hybrid mode cavity has the potential to achieve tight optical confinement with high-quality factors, making it a promising candidate for low-threshold lasers. These lasers exploit enhanced spontaneous emission into the lasing mode, which leads to a high spontaneous emission coupling ratio.[29] To numerically simulate lasing from the hybrid cavity, we focus on the specific case where the active material in the GaAs layer is composed of InAs quantum dots. The quantum dots can be described by a four-level atom model.[25,26] In our numerical simulation, we employ a pump of 750 nm wavelength with a pulse width of 4.5 ps and a broadband probe of two orders of magnitude smaller than the pump amplitude centered at the hybrid cavity resonance wavelength of 1516.53 nm. In the four-level model, we use an initial population volume density of $N_1 = N_0 = 7 \times 10^{21}/m^3$.

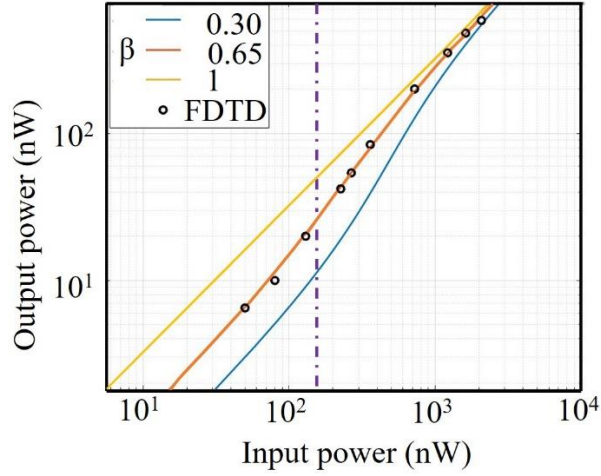

Figure 3. Using 3D FDTD, we calculate the input power versus output power (log-log L-L curve) of the hybrid Si-GaAs cavity nanolaser (black circles) and fit the results using a rate equation model (Equation 1) as shown by the orange curve, which correspond to spontaneous emission coupling ratios of β = 0.65. The yellow and blue solid curves are also plotted using rate equation with spontaneous emission coupling ratios of β = 0.3 and 1, respectively. The vertical dashed line is the lasing threshold obtained from the rate equation fit to the FDTD data.

To investigate the lasing capabilities of the hybrid Si-GaAs cavity, we use 3D FDTD to calculate the cavity output power as a function of the input pump power (L-L curve) using the pump-probe technique. Light enters and is coupled in-plane within the heterogeneous region of the silicon and GaAs with the resulting light transmitting from the other side of the device (Figure 1a). Figure 3 plots the FDTD calculations (black circles), which is fitted by the solid lines using the cavity laser rate-equation model given by[30]

$$P_{in} = \frac{\hbar\omega\gamma}{\beta\eta_{in}} \left[ \frac{p}{1+p}(1+\zeta)(1+\beta p) - \zeta\beta p \right] \qquad (1)$$

Here, $P_{in}$ is the input pump power, ω is the cavity resonance frequency, γ = ω/Q is the cavity decay rate, and β is the spontaneous emission coupling ratio. We define the cavity photon number as p = $P_{out}/\hbar\omega\gamma\eta_{out}$, where $\eta_{out}$ is the output collection efficiency of the laser. Additionally, $\eta_{in}$ is the pumping efficiency, and ζ is the cavity photon number at transparency.

From Equation 1, we calculate several L-L curves corresponding to spontaneous emission coupling ratios and find the FDTD lasing data is best fit by a spontaneous emission coupling ratio of 0.65 (orange curve). We then calculate the lasing threshold power from the condition when there is an average of one photon in the cavity (p = 1). Using this value of p in Equation 1, we obtain a threshold power of $P_{th}$ = 156 nW which is shown by the vertical dashed line. We also plot L-L curves corresponding to spontaneous emission coupling ratios of β = 0.3 (blue curve) and 1 (yellow curve) for reference. A laser with large β decreases the lasing threshold, and for β=1, the laser has no threshold, while decreasing β increases the lasing threshold (e.g., β=0.3). A hallmark feature of lasing is an increase in the slope of the log-log curve near the lasing threshold,[31,32] as we observe in the FDTD data (black circles). Our simulation demonstrates a lower lasing threshold compared to other devices using hybrid Si-III/V materials,[15–19] which have previously reported a lasing threshold power as low as 1 µW.[19]

## 4. Bistability

Optical bistability, a nonlinear optical phenomenon, can be defined as an optical system possessing two different output states corresponding to the same input intensity. The optical bistability of two-level atoms confined in an optical cavity is particularly useful for a wide range of applications, including all-optical switches,[33] memories,[34] optical logic gates,[35] and optical transistors.[36] Our hybrid cavity can operate as a low-power bistable device by employing such optical nonlinearity. Specifically, we use a two-level one electron numerical model to simulate the quantum dots as saturable absorbers in the hybrid Si-GaAs cavity.[37,38] Here we use two-level quantum dots for bistability analysis instead of four-level quantum dots used for lasing calculation in the previous section because it is faster to simulate, and this analysis does not require gain media. To simulate the optical response of the InAs quantum dots in the GaAs region of the hybrid cavity, we incorporate the Maxwell-Bloch equations into numerical FDTD simulations using the method described by Shih-Hui Chang and Allen Taflove.[25] The Maxwell Bloch equations describe the response of saturable absorbers to an incident electric field.

$$\frac{dN_{11}}{dt} = -i\frac{\Omega}{2}(\chi_{SA}^* - \chi_{SA}) + \gamma N_{22} \tag{2}$$

$$\frac{dN_{22}}{dt} = i\frac{\Omega}{2}(\chi_{SA}^* - \chi_{SA}) - \gamma N_{22} \tag{3}$$

$$\frac{d\chi_{SA}}{dt} = -(\beta + i\Delta)\chi_{SA} - i\frac{\Omega}{2}(N_{22} - N_{11}) \tag{4}$$

Here, $N_{11}$ and $N_{22}$ are the population densities of the two-level atoms for the ground and excited states, $\Omega = \mu E/\hbar$ is the optical Rabi frequency, and µ is the transition matrix elements of the two-level atom. We define the detuning frequency as $\Delta = \omega_0 - \omega$, while $\omega_0$ and $\omega$ are the resonant frequency and incident light frequency. We also describe the atomic decay rate as $\gamma = \gamma_{non} + \gamma_{rad}$, where $\gamma_{non}$ is the nonradiative decay rate and $\gamma_{rad} = \frac{\mu^2 \omega^3}{3\pi\hbar\epsilon_0(1+\chi_D)c^3}$ is the radiative decay rate. Finally, $\beta = \gamma/2 + 1/T_2$, while $T_2$ is the dipole dephasing time.

For the simulation parameters, we set a layer of InAs quantum dots in the GaAs region of the hybrid cavity at a planar quantum dot density of N =7×10¹⁰ /cm².[39] We set the decay rates of the quantum dots to their room temperature values of $\gamma_{rad}$ = 1 GHz[40] and $\gamma_{non}$ = 1 GHz,[41] and the dephasing time to T₂ = 300 fs.[42]

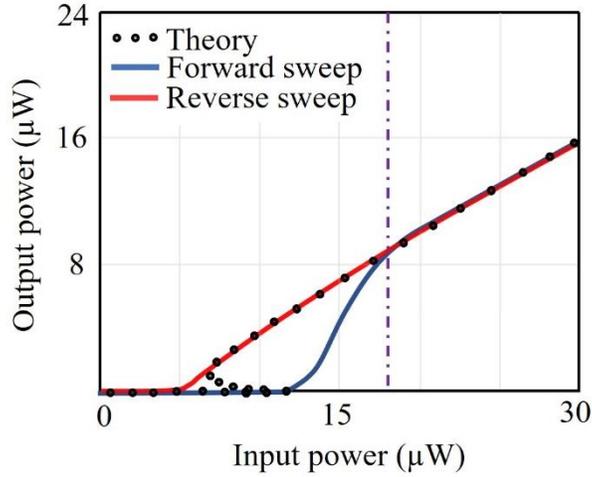

Figure 4. Output power vs. input power at the resonance wavelength 1516.53 nm of the hybrid Si-GaAs cavity, featuring quantum dots as saturable absorbers in the GaAs slab. The blue and red curves are the forward and reverse sweeps of the input intensity, while the black circles represent the theoretical bistability curve. The vertical purple dashed line is the bistability threshold.

Figure 4 plots the on-resonance bistability in the hybrid cavity for in-plane (Figure 1a) continuous wave excitation at 1516.53 nm. Theoretically, the output intensity is a bistable function of the cavity input at a steady state, as shown by the green dashed line in Figure 4 corresponds to the following equation[43,44]

$$y = x + \frac{2Cx}{1+x^2} \quad (5)$$

Where C = αL/2T is the co-operativity parameter, with α being the absorption coefficient, and x and y are the output field amplitude and input field amplitude. We evaluate the co-operativity parameter C = 10.49 from FDTD calculated atomic-absorption coefficient α = $1.37 \times 10^4$ cm$^{-1}$.

However, in FDTD, we observe a hysteretic behavior in the output intensity depending on whether the input intensity is increasing or decreasing (blue and red curves, respectively).[45] Bistable loop of output power provides the bistability threshold power, which occurs at 18.1 µW according to our FDTD calculations (vertical purple dashed line). The low bistability threshold of this hybrid cavity design is promising for various silicon photonic applications at telecom wavelengths requiring low-power optical nonlinearity. Our design provides a comparable threshold as observed in silicon photonic crystal cavities[46-48] and ring cavities.[49-51]

## 5. Conclusion
In summary, we have shown that low power threshold lasing and optical bistability can be achieved using a new Si-GaAs hybrid cavity design. In this approach, the cavity is patterned in the silicon waveguide while the GaAs slab features quantum dots that are optically active at telecom wavelengths. The hybrid device is facile to fabricate without careful alignment using simple pickup and placement by transfer printing method[52,53] and can be practically implemented. Additionally, this hybrid device can operate at different frequencies by changing the cavity design parameters, with optimization enabling further improved performance. Our work represents a critical step toward the efficient heterogeneous integration of silicon and III-V materials for future low-power silicon nanophononics.

**Acknowledgement**
Office of Naval Research (grant #N000142012551), ARL (grant #W911NF1920181), AFOSR-AOARD (grant #FA23862014072), and National Science Foundation (Grants #OMA1936314, #ECCS1933546, #PHY1839165).